\RequirePackage[2020-02-02]{latexrelease}
\documentclass[aps,prl,twocolumn,showpacs]{revtex4}
\usepackage{amssymb}
\usepackage{amsmath}
\usepackage{graphicx}
\usepackage{color}

\begin{document}
\title{Two-photon interference from a quantum emitter in hexagonal boron nitride}
\author{Clarisse Fournier$^{1}$, S\'ebastien Roux$^{1,2}$,  Kenji Watanabe$^3$, Takashi Taniguchi$^4$, St\'ephanie Buil$^{1}$, Julien Barjon$^{1}$, Jean-Pierre Hermier$^1$, Aymeric Delteil$^1$}

\affiliation{$^1$ Universit\'e Paris-Saclay, UVSQ, CNRS,  GEMaC, 78000, Versailles, France. \\
$^2$ Universit\'e Paris-Saclay, ONERA, CNRS, Laboratoire d'\'etude des microstructures, 92322, Ch\^atillon, France. \\
$^3$ Research Center for Functional Materials, 
National Institute for Materials Science, 1-1 Namiki, Tsukuba 305-0044, Japan \\
$^4$ International Center for Materials Nanoarchitectonics, 
National Institute for Materials Science, 1-1 Namiki, Tsukuba 305-0044, Japan }
\affiliation{aymeric.delteil@usvq.fr}


\begin{abstract}

Recently discovered quantum emitters in two-dimensional (2D) materials have opened new perspectives of integrated photonic devices for quantum information. Most of these applications require the emitted photons to be indistinguishable, which has remained elusive in 2D materials. Here, we investigate two-photon interference of a quantum emitter generated in hexagonal boron nitride (hBN) using an electron beam. We measure the correlations of zero-phonon-line photons in a Hong-Ou-Mandel (HOM) interferometer under non-resonant excitation. We find that the emitted photons exhibit a partial indistinguishability of $0.44 \pm 0.11$ in a 3~ns time window, which corresponds to a corrected value of $0.56 \pm 0.11$ after accounting for imperfect emitter purity. The dependence of the HOM visibility on the width of the post-selection time window allows us to estimate the dephasing time of the emitter to be $\sim 1.5$~ns, about half the limit set by spontaneous emission. A visibility above 90~\% is under reach using Purcell effect with up-to-date 2D material photonics. 

\end{abstract}

 \maketitle
Two-photon interference is essential for many photonic implementations of quantum information protocols, from linear optical quantum computing~\cite{klm01} to distant entanglement generation~\cite{Cabrillo99, Bernien13, Delteil15} and quantum communication~\cite{Gisin07}. The indistinguishability of two single-photon pulses -- which quantifies their ability to interfere -- results in the so-called Hong-Ou-Mandel (HOM) effect~\cite{hom87}. The latter refers to the fact that perfectly indistinguishable photons simultaneously reaching the two input ports of a beamsplitter always exit the beamsplitter from the same output port~\cite{Bouchard21}. Experimental observation of the HOM effect between consecutive photons from a quantum emitter constitutes an important milestone in the use of a physical system for the generation of scalable photonic qubits. Among the physical systems able to generate indistinguishable photons, solid-state single-photon emitters (SPEs) have been widely investigated due to their potential for integration in photonic devices~\cite{Aharonovich16}. Thus, photon indistinguishability has been experimentally demonstrated with III-V semiconductor quantum dots~\cite{Santori02, Somaschi16, Thoma17, Gao13} and color centers in three-dimensional wide bandgap crystals~\cite{Sipahigil12, Sipahigil14, Morioka20}.

In turn, recently discovered quantum emitters in 2D materials, comprising trapped excitons in transition metal dichalcogenides~\cite{chakraborty15, he15, koperski15, srivastava15, tonndorf15} and color centers in hBN~\cite{tran16, bourrelier16, martinez16}, have raised a growing interest owing to the perspectives of extreme miniaturization and integration into complex heterostructures~\cite{Geim13} -- yet without demonstration of two-photon interference to date. Among these systems, a recently discovered family of hBN SPEs stands out -- a class of blue emitting color centers (abbreviated B-centers in the following) that can be generated at controlled locations using an electron beam. Their zero-phonon-line (ZPL) center wavelength is consistently found within 3~meV around 436~nm~\cite{shevitski19, Fournier21,Gale22}. Several studies have already demonstrated their spectral stability, narrow linewidth, brightness and single-photon emission up to room temperature~\cite{Fournier21, Gale22, Horder22}.

\begin{figure}[h]
  \centering
  \includegraphics[width=3.5in]{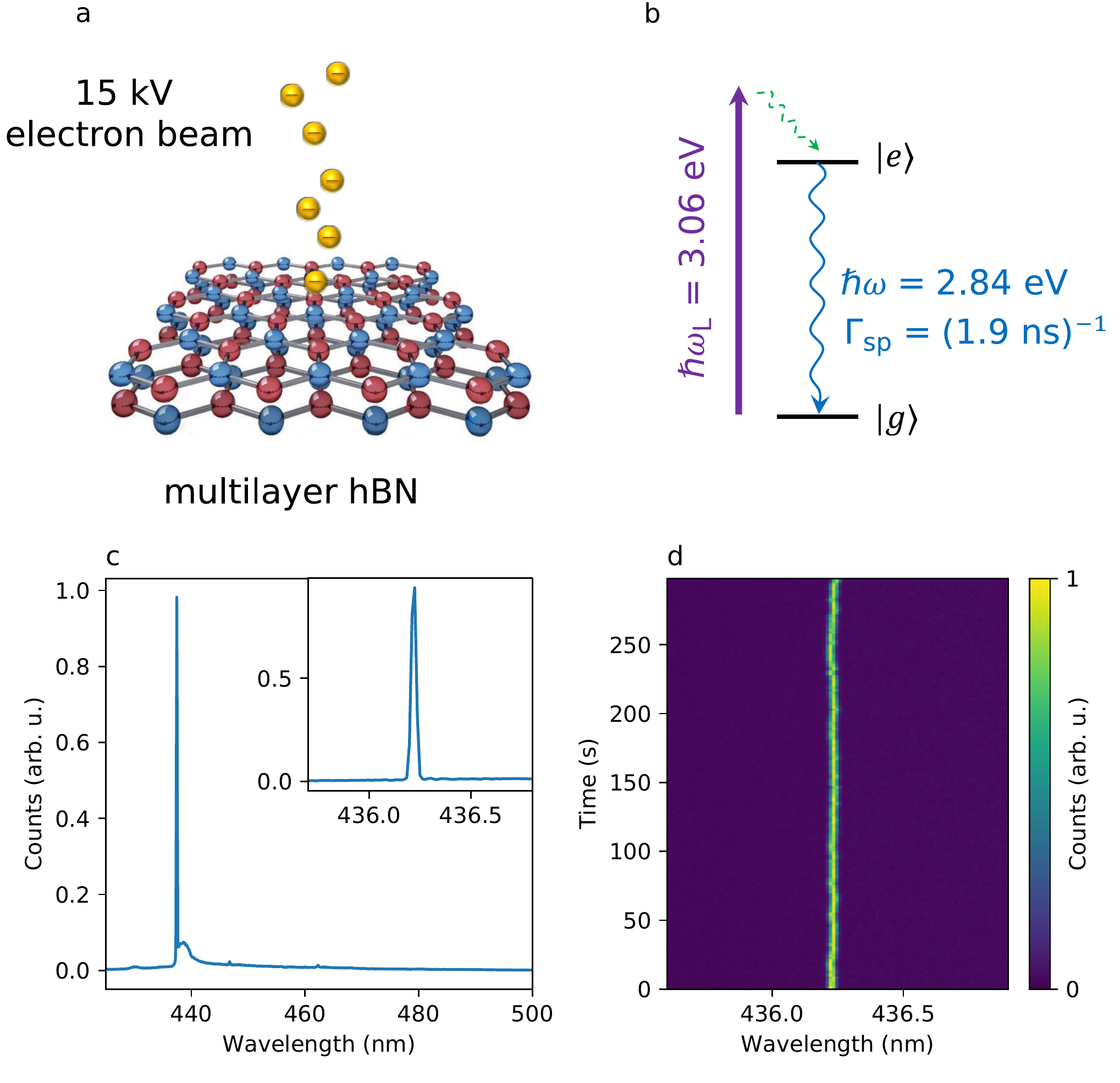}\\
  \caption{(a) Irradiation by a 15~kV electron beam generates B-centers in a multilayer hBN crystal. (b) Energy levels of the SPE: A 405~nm laser excites the emitter. Non-radiative relaxation occurs, followed by emission of a photon at 436~nm. (c) Low resolution spectrum of the SPE, where the ZPL and the acoustic phonon sideband can be observed. Inset: High resolution spectrum, limited by the spectrometer resolution of about 100~$\mu$eV. The phonon pedestal of the main panel is no more visible due to the higher resolution. (d) High resolution spectra as a function of time measured with 2~s integration time during 5~min, showing the stability of the SPE.}\label{figure1}
\end{figure}

In this letter, we characterize two-photon interference of light emitted by an individual B-center. Our sample consists of a single hBN crystal grown using high pressure, high temperature conditions~\cite{Taniguchi07}, that we exfoliated on a SiO$_2$(285~nm)/Si substrate. We generate a SPE ensemble in a commercial scanning electron microscope (SEM) using a slightly defocused electron beam (diameter $\sim$300~nm) under 15~kV acceleration voltage and 10~nA current, following~\cite{Fournier21} (figure~\ref{figure1}a). We subsequently characterize the sample in a confocal microscope operating in a helium closed-cycle cryostat, keeping the sample at 4~K. The sample is optically excited by a pulsed diode laser of 405~nm wavelength, 850~$\mu$W power and 80~MHz repetition rate (figure~\ref{figure1}b) that is focused on the sample using a microscope objective of numerical aperture~0.8. The SPE luminescence is collected through the same objective, and coupled into a single-mode fiber. In the following, we focus on an individual SPE with a ZPL centered at 436.24~nm. Figure~\ref{figure1}c shows a low resolution spectrum of the SPE, exhibiting the usual spectral shape of the B-centers, which comprises a narrow ZPL (40~\% of the emission) and an acoustic phonon sideband (60~\%). We ensure that spectral diffusion of the ZPL is limited, as shown figure~\ref{figure1}d, where the wavelength fluctuations are contained below 20~pm.

\begin{figure*}[ht]
  \centering
  \includegraphics[width=7in]{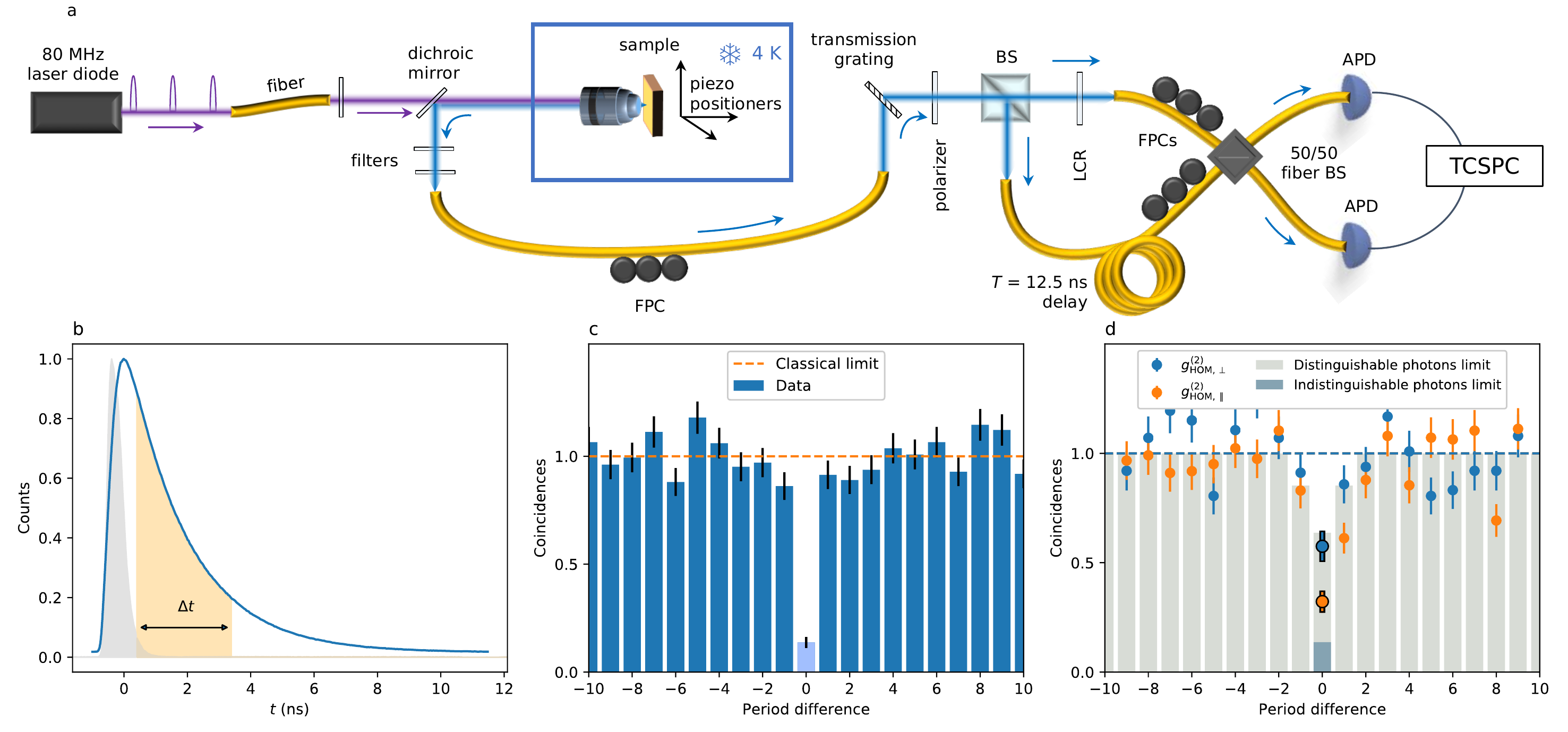}\\
  \caption{(a) Experimental setup for the HOM experiment. LCR: Liquid crystal retarder; BS: Beamsplitter; APD: Avalanche photodiode; FPC: Fiber polarization controller. TCSPC: time-correlated single photon counter. (b) Histogram of the photon detection events at the output port. The instrument response function is indicated by the gray shadow. The orange shading of width $\Delta t = 3$~ns indicates the post-selection time window. A fit to the data (not shown) provides the spontaneous emission time $\Gamma_\mathrm{sp}^{-1} = 1.9$~ns. (c) Second-order correlation function of the photons emitted during $\Delta t$, yielding $g^{(2)}(0) = 0.14 \pm 0.03$. The horizontal axis is in units of repetition periods. (d) Two-photon coincidences in Hong-Ou-Mandel configuration. The orange (blue) dots denote the normalized coincidence rate for parallel (orthogonal) polarization configuration of the second beamsplitter input ports. The center peak provides the values $g^{(2)}_\mathrm{HOM, \parallel}(0) = 0.32 \pm 0.05$ (parallel case) and $g^{(2)}_\mathrm{HOM, \perp}(0) = 0.58 \pm 0.07$ (orthogonal case). The light gray bars marks the theoretical values for distinguishable photons of $g^{(2)}(0) = 0.14$. The dark gray bar indicates the theoretical value for fully indistinguishable photons of $g^{(2)}(0) = 0.14$.}\label{figure2}
\end{figure*}

Figure~\ref{figure2}a depicts the experimental setup used for two-photon interference characterization. The photoluminescence is collected in a single-mode fiber that channels the photons to a delayed Mach-Zehnder interferometer. The ZPL is filtered using a transmission grating of 1379~grooves per millimeter. Together with subsequent coupling to single-mode fibers, it implements a narrow bandpass filter of 100~GHz bandwidth. This spectral width is larger than the time-averaged linewidth but much narrower than the width of the acoustic phonon sideband of about 7.8~meV (1.9~THz). A polarizer ensures that the input photons have a well-defined linear polarization at the input port of the delayed Mach-Zehnder interferometer. One of the arms is delayed by the same amount as our repetition period (12.5~ns) using a 2.6~m fiber, such that two consecutively emitted photons can simultaneously impinge on the beamsplitter. A liquid crystal retarder is inserted in the other arm to rotate the photon polarization by 90~degrees when suited, allowing photon polarization at the two input ports of the second beamsplitter to be either identical (parallel) or orthogonal. The total count rate of the ZPL photons at the output ports is about 1200 counts per second. Figure~\ref{figure2}b shows a histogram of the photon detection times at the output ports. We first consider photon detections occurring during the $\Delta t =$~3~ns time window highlighted by the orange shadowing on figure~\ref{figure2}b, that is located after the laser pulse of width 550~ps (gray shadowing on figure~\ref{figure2}b). Figure~\ref{figure2}c shows the second-order photon correlations measured in a Hanbury Brown and Twiss configuration of the interferometer (\textit{i.e.} in a single arm). The relative height of the center period allows to infer the photon purity of $g^{(2)}_\mathrm{HBT}(0) = 0.14 \pm 0.03$, limited by reminiscent background signal and dark counts.

We then measure the photon coincidences in the HOM configuration of the interferometer. Figure~\ref{figure2}d gives the normalized coincidences measured during 36~hours while alternating between parallel and orthogonal polarizations using the liquid crystal retarder, considering photons detected during the same time window of width $\Delta t$. The significant reduction of the center period value $g^{(2)}_\mathrm{HOM, \parallel}(0) = 0.32 \pm 0.05$ in the parallel polarization case as compared with the orthogonal case  $g^{(2)}_\mathrm{HOM, \perp}(0) = 0.58 \pm 0.07$ is a signature of photon coalescence. The raw (uncorrected) degree of indistinguishability of the emitted photons is then given by the interference visibility defined as $V_\mathrm{HOM} = 1 - g^{(2)}_\mathrm{HOM, \parallel}(0) / g^{(2)}_\mathrm{HOM, \perp}(0)$. In our case, we find  $V_\mathrm{HOM}= 0.44 \pm 0.11$. In the case of single photons with ideal purity (\textit{i.e.} $g^{(2)}_\mathrm{HBT}(0) = 0$), this quantity ranges between $V_\mathrm{HOM} = 0$ (perfectly distinguishable photons) and $V_\mathrm{HOM} = 1$ (perfectly indistinguishable photons). When  multiple detection events are not negligible, the theoretical bounds of $g^{(2)}_\mathrm{HOM, \parallel}(0)$ and $g^{(2)}_\mathrm{HOM, \perp}(0)$ are offset upwards~\cite{Santori02}, such that the corrected visibility reads $V_\mathrm{HOM}^\mathrm{corr} = \left( 1 + 2 g^{(2)}_\mathrm{HBT}(0) \right) V_\mathrm{HOM}$. In the case of the studied SPE, the theoretical upper (resp. lower) bounds accounting for our finite value of $g^{(2)}_\mathrm{HBT}(0)$ are indicated by light (resp. dark) gray bars on the center period of the histogram shown figure~\ref{figure2}d. Accordingly, we find a corrected HOM visibility of $V_\mathrm{HOM}^\mathrm{corr} = 0.56 \pm 0.11$. This number is comparable with the indistinguishability of non-resonantly excited quantum dots~\cite{Santori02, Thoma17} or color centers in wide gap 3D semiconductors~\cite{Morioka20}. This observation of HOM interference from single photons emitted by a 2D material quantum emitter constitutes the main result of our study, and suggests possible practical applications of hBN for optical quantum information, upon further improvement of the visibility.

\begin{figure}[t]
  \centering
  \includegraphics[width=3.5in]{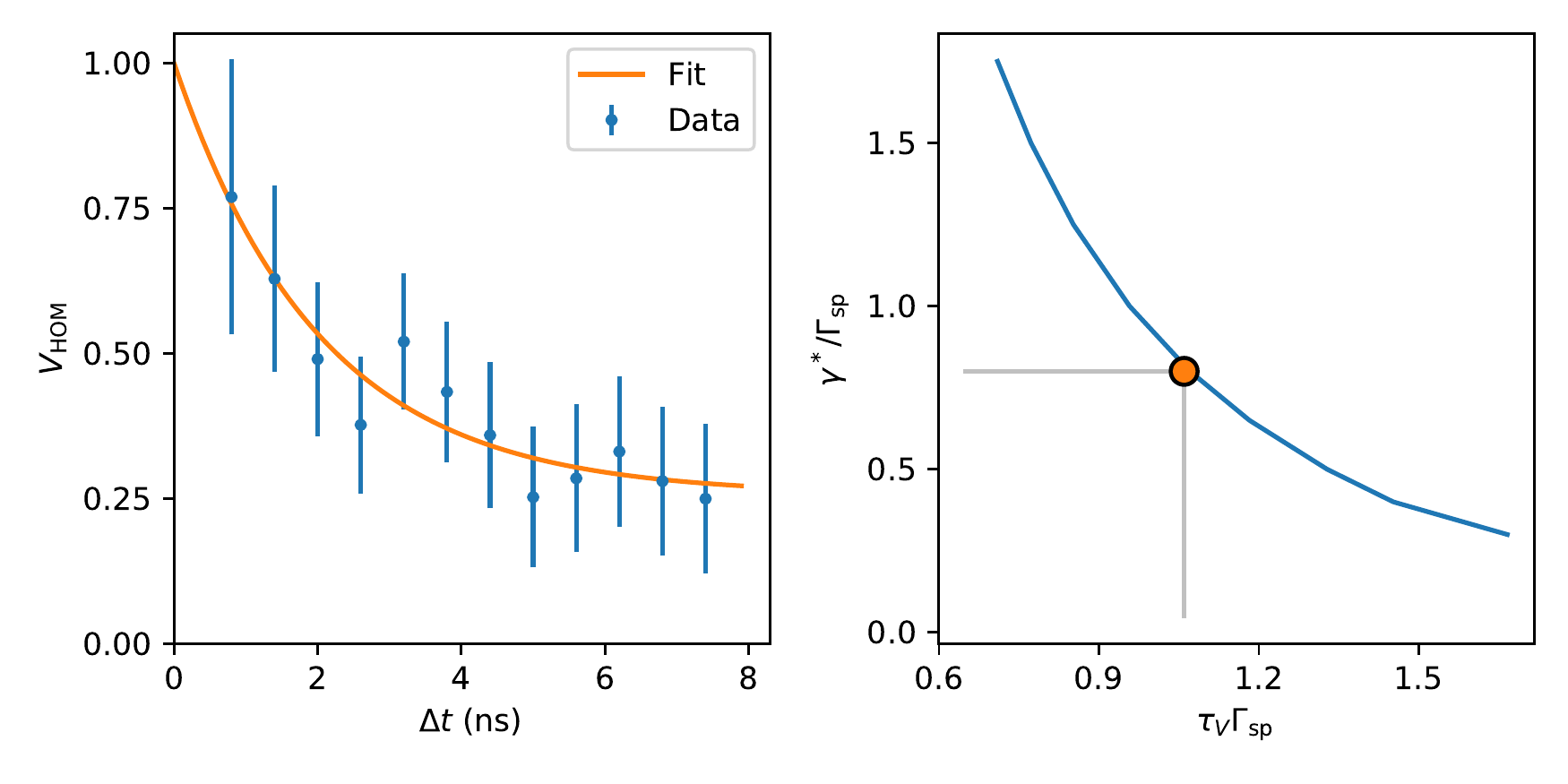}\\
  \caption{(a) Blue dots: HOM visibility $V_\mathrm{HOM}^\mathrm{corr}$ as a function of the post-selection time window width $\Delta t$ . Orange line: exponential fit to the data, yielding $\tau_V = 1.7$~ns. (b) Blue line: Simulated value of $\tau_V$ as a function of the dephasing rate $\gamma^*$, in units of radiative lifetime $\Gamma_\mathrm{sp}^{-1}$. Orange dot: Value corresponding to the experimental data (a). }\label{figure3}
\end{figure}

The limited value of the corrected HOM visibility can be possibly attributed to fast dephasing of the optical dipole. The total dephasing rate of an optical transition can be written $\gamma = \Gamma_\mathrm{sp}/2 + \gamma^*$, where $\Gamma_\mathrm{sp} = 1/T_1$ is the spontaneous emission rate, and $\gamma^* = 1/T_2^*$ denotes the rate of dephasing caused by reservoirs other than the vacuum electromagnetic field. In the pulsed regime, only $\gamma^*$ causes a reduction of the photon indistinguishability~\cite{kiraz04, bylander03}. An expected consequence of dephasing is that extending the integration window $\Delta t$ would degrade the HOM visibility by allowing a larger delay time between detected pairs~\cite{Bouchard21, Martinez22, Proux15,Kim16,Nawrath19}. It is then in principle possible to estimate $\gamma^*$ by observing the influence of the postselection time window on the interference visibility. Figure~\ref{figure3}a (blue dots) plots the measured HOM visibility $V_\mathrm{HOM}^\mathrm{corr}$ as a function of the post-selection window size $\Delta t$. We can observe an overall decrease of the HOM visibility as $\Delta t$ increases, which is consistent with the effect of dephasing. The decay of $V_\mathrm{HOM}^\mathrm{corr}(\Delta t)$ can be fitted by an exponential function, of which we have fixed the intercept to be~1. This result is compatible with the assumption that finite purity and dephasing are the main sources of imperfect HOM visibility. The fit yields the decay time $\tau_V = 2.0$~ns. To relate this timescale to an estimation of $\gamma^*$, we simulate the two-photon interference visibility of a two-level atom with dephasing based on numerical integration of a master equation~\cite{loudon, Fischer16, qutip1, qutip2} accounting for spontaneous emission and dephasing in the Lindblad form. Our simulation accounts for a time post-selection of the coincidences in the same way as experimentally realized. The details of the simulation are exposed in the Supplemental Material~\cite{supplemental}. We compute the two-photon interference visibility while varying the post-selection time window, from which we extract the visibility decay timescale $\tau_V$. By repeating this process for different values of the dephasing rate $\gamma^*$, we obtain a one-to-one relation between $\tau_V$ and $\gamma^*$, which we plot on figure~\ref{figure3}b. The dephasing rate associated with our experimental data corresponds to $T_2^* = 2.4 \pm 0.7$~ns. This value is in agreement with prior estimations of B-center dephasing time based on the decay of Rabi oscillations in resonant excitation~\cite{Horder22}, where values of $T_2^*$ between 0.7 and more than 2~ns were estimated, depending on the emitter and on the laser power. In turn, this places the total dephasing time $T_2 = (\Gamma_\mathrm{sp}/2 + \gamma^*)^{-1} = 1.5$~ns, about halfway from the perfect indistinguishability limit $T_2 = 2 T_1 = 3.8$~ns.

The value of indistinguishability we measured could be further improved by using  resonant excitation, which selectively addresses the transition of interest. This leads to a decrease of the saturation power by several orders of magnitude and therefore of the environment noise, yielding up to Fourier-limited linewidths~\cite{Fournier23}. Additionally, embedding the emitter in a cavity would increase the ratio between the rates of spontaneous emission and dephasing. We calculate that a Purcell factor of 7 (15) would increase the indistinguishability to 80~\% (90~\%), while improving at the same time the collection efficiency. Similar or higher spontaneous emission enhancements have already been achieved in 2D materials~\cite{Froch21, Iff21}. Integrating B-centers into electrical gates for fine tuning of the optical resonance would then allow interference between distinct emitters~\cite{Zhigulin23}. The indistinguishability of photons emitted by hBN B-centers, together with their already-established advantageous photophysical properties and the possibility to locate them at a prechosen position, opens the way to a broad range of applications in integrated quantum photonics and optical quantum information based on 2D materials.

\section{Acknowledgments}
 The authors acknowledge Aur\'elie Pierret and Michael Rosticher for the flake exfoliation. This work is supported by the French Agence Nationale de la Recherche (ANR) under reference ANR-21-CE47-0004-01 (E$-$SCAPE project). This work also received funding from the European Union’s Horizon 2020 research and innovation program under Grant No. 881603 (Graphene Flagship Core 3). K.W. and T.T. acknowledge support from JSPS KAKENHI (Grant Numbers 19H05790, 20H00354 and 21H05233).


\pagebreak
~
\newpage

\onecolumngrid
\begin{center}
  \textbf{\large Supplementary Material\\~\\Two-photon interference from a quantum emitter in hexagonal boron nitride}\\[.2cm]
  Clarisse Fournier$^{1}$, S\'ebastien Roux$^{1,2}$, Kenji Watanabe$^3$, Takashi Taniguchi$^4$, Julien Barjon$^{1}$, St\'ephanie Buil$^{1}$, Jean-Pierre Hermier$^1$, Aymeric Delteil$^1$\\[.1cm]
  {\itshape \small $^1$ Universit\'e Paris-Saclay, UVSQ, CNRS,  GEMaC, 78000, Versailles, France. \\
  $^2$ Universit\'e Paris-Saclay, ONERA, CNRS, Laboratoire d'\'etude des microstructures, 92322, Ch\^atillon, France. \\
$^3$ Research Center for Functional Materials, 
National Institute for Materials Science, 1-1 Namiki, Tsukuba 305-0044, Japan \\
$^4$ International Center for Materials Nanoarchitectonics, 
National Institute for Materials Science, 1-1 Namiki, Tsukuba 305-0044, Japan \\
{\color{white}--------------------} aymeric.delteil@usvq.fr{\color{white}--------------------} \\[1cm]}

\end{center}

\setcounter{equation}{0}
\setcounter{figure}{0}
\setcounter{table}{0}
\setcounter{page}{1}
\renewcommand{\theequation}{S\arabic{equation}}
\renewcommand{\thefigure}{S\arabic{figure}}
\renewcommand{\bibnumfmt}[1]{[S#1]}
\renewcommand{\citenumfont}[1]{S#1}

\textbf{Influence of the post-selection on the HOM visibility} \\

It is well known that dephasing degrades the visibility of two-photon interference when the time difference between coincidences is larger than the dephasing time~[33-35]. Time postselection of photon detection events, allowed by suitable detector resolution, can thus increase the Hong-Ou-Mandel (HOM) visibility and has been used to boost the fidelity of quantum protocols based on HOM interference -- at the price of a lower count rate~[3,4]. In this section, we calculate the evolution of the HOM visibility upon modification of the post-selection time window based on numerical integration of the master equation.

Our system is modeled by a two-level system with computational states $|g\rangle$ and $|e\rangle$, and is described in the rotating frame, in the absence of coherent laser drive. As a consequence, the time evolution is purely dissipative, governed by the Lindblad master equation $\dot{\rho} = \sum_{c_i} c_i \rho c_i^\dagger - \frac{1}{2}\left\lbrace c_i^\dagger c_i, \rho \right\rbrace$, with $c_i$ the relevant collapse operators. The dissipation terms are spontaneous emission at rate $\Gamma_\mathrm{sp}$ and pure dephasing at rate~$\gamma^*$. The corresponding collapse operators read, respectively, $\sqrt{\Gamma_\mathrm{sp}} \sigma^-$ and $\sqrt{\gamma^*/2}\sigma_z$ (in the convention where $\gamma^* = 1/T_2^*$~[32]), where $\sigma_z =|e\rangle \langle e | - |g\rangle \langle g |$, and $\sigma^- = |g \rangle \langle e |$. The effect of the non-resonant laser pulse is accounted for by taking $|e\rangle$ as the initial state. In the far field, the field operators are locked to the source operators~[36], such that $a(t) \propto \sigma^-\left( t - r/c \right)$, with $a(t)$ the photon annihilation operator. The field observables are therefore expressed in terms of source operators in the following.

The HOM visibility can be expressed as a function of the first- and second-order coherence of the light field~[37]:
\[
g^{(2)}_\mathrm{HOM}(0) = \frac{1}{2} g^{(2)}_\mathrm{HBT}(0) + \frac{1}{2} \left(
1 - \frac{1}{N^2} \int \int \mathrm{d}t  \mathrm{d}t ' \left| G^{(1)}(t, t') \right|^2 
\right)
\]
with $G^{(1)}(t, t') = \langle \sigma^+(t) \sigma^-(t') \rangle$, $N$ a normalization factor $N = \int \mathrm{d}t \langle \sigma^+(t) \sigma^- (t) \rangle = \int \mathrm{d}t \langle \sigma_{ee} (t)\rangle$ and $g^{(2)}_\mathrm{HBT}(0) = 0$ for an ideal single-photon source.

 $ G^{(1)}(t, t')$ is numerically calculated based on the quantum regression theorem using the Qutip toolbox~[38,39]. To account for the temporal post-selection, the integrals run over $t, t' \in [0, \Delta t]$, where $\Delta t$ is the post-selection window size. $g^{(2)}_\mathrm{HOM}(0)$ is calculated as a function of $ \Delta t$ for various values of $\gamma^*$.
 
\begin{figure*}[h!] 
\centering
\includegraphics[width=0.6\textwidth]{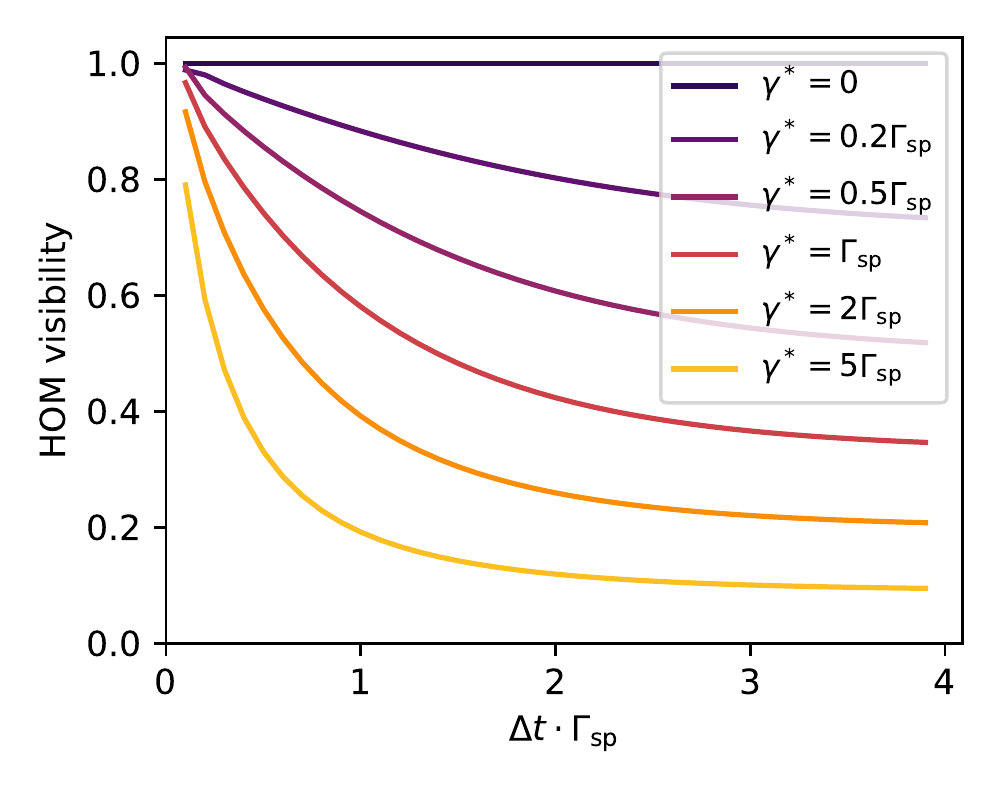}
\caption{$V_\mathrm{HOM}$ as a function of $\Delta t$ for different values of $\gamma^*$.    } \label{figS1}
\end{figure*}

Figure~\ref{figS1} shows the result of the calculation for six values of $\gamma^*$. It can be seen that the visibility $V_\mathrm{HOM} = 1 - 2 g^{(2)}_\mathrm{HOM}(0) $ decays with increasing $\Delta t$. The decay time depends on the dephasing, and is plotted as a function of $\gamma^*$ in main text figure~3b. The visibility for $\Delta t \approx 0$ is always high since the photons are always indistinguishable at short delays~[33], and decays faster for larger dephasing rates. The long-time asymptote, corresponding to integration of all detection events, coincides with the bare (non-postselected) indistinguishability $T_2/2T_1 = \Gamma_\mathrm{sp}/(\Gamma_\mathrm{sp} + 2 \gamma^*)$ [30].
\vspace{1cm}

\newpage
\vspace{1cm}
\textbf{ Influence of the Purcell factor on the HOM visibility} \\

As mentioned in the previous section, the integrated HOM visibility (without any post-selection) for an ideal emitter is $V_\mathrm{HOM} = T_2/2T_1$ [30] -- the ideal case $T_2 = 2 T_1$ providing $V_\mathrm{HOM} = 1$. In a cavity, the Purcell effect modifies the lifetime according to $T_1' = T_1/F_p$, where $F_p$ is the Purcell factor. Given our estimation of $T_2^* = 2.4$~ns, an indistinguishability of 0.80 (resp 0.90) corresponds to $F_p = 7$ (resp. $F_p = 15$).

%
%
%



\begin{thebibliography}{99}

\bibitem{klm01} E. Knill, R. Laflamme, and G. J. Milburn, A scheme for efficient quantum computation with linear optics, Nature \textbf{409}, 46 (2001).

\bibitem{Cabrillo99} C. Cabrillo, J. I. Cirac, P. Garc\'ia-Fern\'andez, and P. Zoller, Creation of entangled states of distant atoms by interference, Phys. Rev. A \textbf{59}, 1025 (1999).

\bibitem{Bernien13} 
H. Bernien, B. Hensen, W. Pfaff, G. Koolstra, M. S. Blok, L. Robledo, T. H. Taminiau, M. Markham, D. J. Twitchen, L. Childress, and R. Hanson, Heralded entanglement between solid-state qubits separated by three metres, Nature \textbf{497}, 86 (2013).

\bibitem{Delteil15} A. Delteil, Z. Sun, W.-B. Gao, E. Togan, S. F\"alt, and A. Imamo\u{g}lu, Generation of heralded entanglement between distant hole spins,  Nature Phys. \textbf{12}, 218 (2016).

\bibitem{Gisin07} N. Gisin and R. Thew, Quantum communication, Nat. Photon. \textbf{1}, 165 (2007).

\bibitem{hom87} C. K. Hong, Z. Y. Ou, and L. Mandel, Measurement of subpicosecond time intervals between two photons by interference, Phys. Rev. Lett. \textbf{59}, 2044 (1987).

\bibitem{Bouchard21} F. Bouchard, A. Sit, Y. Zhang, R. Fickler, F. M. Miatto, Y. Yao, F. Sciarrino, and E. Karimi, Two-photon interference: the Hong–Ou–Mandel effect, Rep. Prog. Phys. \textbf{84} 012402 (2021).

\bibitem{Aharonovich16} I. Aharonovich, D. Englund, and M. Toth, Solid-state single-photon emitters, Nat. Photon. \textbf{10} 631 (2016).

\bibitem{Santori02} C. Santori, D. Fattal, J. Vu\v{c}kovi\'c, G. S. Solomon, and Y. Yamamoto, Indistinguishable photons from a single-photon device, Nature \textbf{419}, 594 (2002).


\bibitem{Somaschi16} N. Somaschi, V. Giesz, L. De Santis, J. C. Loredo, M. P. Almeida, G. Hornecker, S. L. Portalupi, T. Grange, C. Ant\'on, J. Demory \textit{et al.}, Near-optimal single-photon sources in the solid state, Nat. Photonics \textbf{10}, 340 (2016).

\bibitem{Thoma17} A. Thoma, P. Schnauber, J. B\"ohm, M. Gschrey, J.-H. Schulze, A. Strittmatter, S. Rodt, T. Heindel, and S. Reitzenstein, Two-photon interference from remote deterministic quantum dot microlenses, Appl. Phys. Lett. \textbf{110}, 011104 (2017).

\bibitem{Gao13} W.-B. Gao, P. Fallahi, E. Togan, A. Delteil, Y.S. Chin, J. Miguel-Sanchez, and A. Imamo\u{g}lu, Quantum teleportation from a propagating photon to a solid-state spin qubit, Nat. Commun. \textbf{4}, 2744 (2013).

\bibitem{Sipahigil12} A. Sipahigil, M. L. Goldman, E. Togan, Y. Chu, M. Markham, D. J. Twitchen, A. S. Zibrov, A. Kubanek, and M. D. Lukin, Quantum Interference of Single Photons from Remote Nitrogen-Vacancy Centers in Diamond, Phys. Rev. Lett. \textbf{108}, 143601 (2012).

\bibitem{Sipahigil14} A. Sipahigil, K. D. Jahnke, L. J. Rogers, T. Teraji, J. Isoya, A. S. Zibrov, F. Jelezko, and M. D. Lukin, Indistinguishable Photons from Separated Silicon-Vacancy Centers in Diamond, Phys. Rev. Lett. \textbf{113}, 113602 (2014).

\bibitem{Morioka20} N. Morioka, N. Morioka, C. Babin, R. Nagy, I. Gediz, E. Hesselmeier, D. Liu, M. Joliffe, M. Niethammer, D. Dasari, V. Vorobyov, \textit{et al.}, Spin-controlled generation of indistinguishable and distinguishable photons from silicon vacancy centres in silicon carbide, Nature Commun. \textbf{11}, 2516 (2020).

\bibitem{chakraborty15}
C. Chakraborty, L. Kinnischtzke, K. M. Goodfellow, R. Beams, and A. N. Vamivakas, Voltage-controlled quantum light from an atomically thin semiconductor, Nat. Nanotechnol. \textbf{10}, 507 (2015).

\bibitem{he15}
Y.-M. He, G. Clark, J. R. Schaibley, Y. He, M.-C. Chen, Y.-J. Wei,
X. Ding, Q. Zhang, W. Yao, X. Xu, Single quantum emitters in monolayer
semiconductors, \textit{et al}, Nat. Nanotechnol. \textbf{10}, 497 (2015).

\bibitem{koperski15} M. Koperski, K. Nogajewski, A. Arora, V. Cherkez, P. Mallet, J.-Y. Veuillen, J. Marcus, P. Kossacki, and M. Potemski, Single photon emitters in exfoliated WSe2 structures
, Nat. Nanotechnol. \textbf{10}, 503 (2015).

\bibitem{srivastava15} A. Srivastava, M. Sidler, A. V. Allain, D. S. Lembke, A. Kis, and A. Imamo\u{g}lu, Optically active quantum dots in monolayer WSe2 Nat. Nanotechnol. \textbf{10}, 491 (2015).

\bibitem{tonndorf15} P. Tonndorf, R. Schmidt, R. Schneider, J. Kern, M. Buscema, G. A. Steele, A. Castellanos-Gomez, H. S. J. van der Zant, S; Michaelis de Vasconcellos, and R. Bratschitsch, Single-photon emission from localized excitons in an atomically thin semiconductor, Optica \textbf{2}, 347 (2015).

\bibitem{tran16} T. T. Tran, K. Bray, M. J. Ford, M. Toth, and I. Aharonovich, Quantum emission from hexagonal boron nitride monolayers, Nat. Nanotechnol. \textbf{11}, 37 (2016).

\bibitem{bourrelier16} R. Bourrelier, S. Meuret, A. Tararan, O. St\'ephan, M. Kociak, L. H. G. Tizei, and A. Zobelli, Bright UV Single Photon Emission at Point Defects in h-BN, Nano Lett. \textbf{16}, 4317 (2016).

\bibitem{martinez16} L. J. Mart\'inez, T. Pelini, V. Waselowski, J. R. Maze, B. Gil, G. Cassabois, and V. Jacques, Efficient single photon emission from a high-purity hexagonal boron nitride crystal, Phys. Rev. B \textbf{94}, 121405 (2016).

\bibitem{Geim13} A. K. Geim, and I. V. Grigorieva, Van der Waals heterostructures, Nature \textbf{499}, 419 (2013).

\bibitem{Fournier21} C. Fournier, A. Plaud, S. Roux, A. Pierret, M. Rosticher, K. Watanabe, T. Taniguchi, S. Buil, X. Quélin, J. Barjon, J.-P. Hermier, A. Delteil, Position-controlled quantum emitters with reproducible emission wavelength in hexagonal boron nitride, Nat. Commun. \textbf{12}, 3779 (2021).

\bibitem{Gale22} A. Gale, C. Li, Y. Chen, K. Watanabe, T. Taniguchi, I. Aharonovich, and M. Toth, Site-Specific Fabrication of Blue Quantum Emitters in Hexagonal Boron Nitride, ACS Photonics \textbf{9}, 2170 (2022).

\bibitem{shevitski19} B. Shevitski, M. Gilbert, C. T. Chen, C. Kastl, E. S. Barnard, E. Wong, D. F. Ogletree, K. Watanabe, T. Taniguchi, A. Zettl, and S. Aloni, Phys. Rev. B \textbf{100}, 155419 (2019). 

\bibitem{Horder22} J. Horder, S. White, A. Gale, C. Li, K. Watanabe, T. Taniguchi, M. Kianinia, I. Aharonovich, and M. Toth, Coherence Properties of Electron-Beam-Activated Emitters in Hexagonal Boron Nitride Under Resonant Excitation, Phys. Rev. Appl. \textbf{18}, 064021 (2022).


\bibitem{Taniguchi07} T. Taniguchi, and K. Watanabe, Synthesis of High-Purity Boron Nitride Single Crystals under High Pressure by using Ba–BN Solvent, J. Cryst. Growth \textbf{303}, 525 (2007).
%
%
%
%
%
%
%
%

\bibitem{bylander03} J. Bylander, I. Robert-Philip, and I. Abram, Interference and correlation of two independent photons, Eur. Phys. J. D \textbf{22}, 295 (2003). 

\bibitem{kiraz04} A. Kiraz, M. Atat\"ure, and A. Imamo\u{g}lu, Quantum-dot single-photon sources: Prospects for applications in linear optics quantum-information processing, Phys. Rev. A \textbf{69}, 032305 (2004).

\bibitem{Martinez22} J. A. Mart\'inez, R. A. Parker, K. C. Chen, . M. Purser, L. Li, C. P. Michaels, A. M. Stramma, R. Debroux, I. B. Harris, M. H. Appel, \textit{et al.}, Photonic Indistinguishability of the Tin-Vacancy Center in Nanostructured Diamond, Phys. Rev. Lett. \textbf{129}, 173603 (2022).

\bibitem{Proux15} R. Proux, M. Maragkou, E. Baudin, C. Voisin, P. Roussignol, and C. Diederichs, Measuring the Photon Coalescence Time Window in the Continuous-Wave Regime for Resonantly Driven Semiconductor Quantum Dots, Phys. Rev. Lett. \textbf{114}, 067401 (2015).

\bibitem{Kim16} J.-H. Kim, T. Cai, C. J. K. Richardson, R. P. Leavitt, and E. Waks, Two-photon interference from a bright single-photon source at telecom wavelengths, Optica \textbf{3}, 577 (2016).

\bibitem{Nawrath19} C. Nawrath, F. Olbrich, M. Paul, S. L. Portalupi, M. Jetter, and P. Michler, Coherence and indistinguishability of highly pure single photons from non-resonantly and resonantly excited telecom C-band quantum dots, Appl. Phys. Lett. \textbf{115}, 023103 (2019).



\bibitem{loudon} R. Loudon, The Quantum Theory of Light, 2nd ed. (Oxford
University Press, New York, 1983).

\bibitem{Fischer16} K. A. Fischer, K. M\"uller, K. G. Lagoudakis, and J. Vu\v{c}kovi\'c, Dynamical modeling of pulsed two-photon interference, New J. Phys. \textbf{18}, 113053 (2016).

\bibitem{qutip1} J. R. Johansson, P. D. Nation, and F. Nori, QuTiP: An opensource Python framework for the dynamics of open quantum systems, Comput. Phys. Commun. \textbf{183}, 1760 (2012).

\bibitem{qutip2} J. R. Johansson, P. D. Nation, and F. Nori, QuTiP 2: A Python framework for the dynamics of open quantum systems, Comput. Phys. Commun. \textbf{184}, 1234 (2013).


\bibitem{supplemental} See Supplemental Material at [URL
  will be inserted by publisher] for the detail of the simulations of time-gated two-photon interference.

\bibitem{Fournier23} C. Fournier, K. Watanabe, T. Taniguchi, S. Buil, J. Barjon, J.-P. Hermier, and A. Delteil, Investigating the fast spectral diffusion of a quantum emitter in hBN using resonant excitation and photon correlations, arXiv:2303.05315 (2023).

\bibitem{Froch21} J. E. Fr\"och, C. Li, Y. Chen, M. Toth, M. Kianinia, S. Kim, and I. Aharonovich, Purcell Enhancement of a Cavity-Coupled Emitter in Hexagonal Boron Nitride, Small \textbf{18}, 2104805 (2021).

\bibitem{Iff21} O. Iff, Q. Buchinger, M. Moczała-Dusanowska, M. Kamp, S. Betzold, M. Davanco, K. Srinivasan, S. Tongay, C. Ant\'n-Solanas, S. H\"ofling, and C. Schneider, Purcell-Enhanced Single Photon Source Based on a Deterministically Placed WSe2 Monolayer Quantum Dot in a Circular Bragg Grating Cavity, Nano Lett. \textbf{21}, 4715 (2021).



\bibitem{Zhigulin23} I. Zhigulin, J. Horder, V. Ivady, S. J. U. White, A. Gale, C. Li, C. J. Lobo, M. Toth, I. Aharonovich, and M. Kianinia, Stark effect of quantum blue emitters in hBN, arXiv:2208.00600 (2022).

 
\end{thebibliography}

\begin{thebibliography}{1}

[3] H. Bernien, B. Hensen, W. Pfaff, G. Koolstra, M. S. Blok, L. Robledo, T. H. Taminiau, M. Markham, D. J. Twitchen, L. Childress, and R. Hanson, Heralded entanglement between solid-state qubits separated by three metres, Nature \textbf{497}, 86 (2013).\\

[4] A. Delteil, Z. Sun, W.-B. Gao, E. Togan, S. F\"alt, and A. Imamo\u{g}lu, Generation of heralded entanglement between distant hole spins,  Nature Phys. \textbf{12}, 218 (2016).\\

[30] J. Bylander, I. Robert-Philip, and I. Abram, Interference and correlation of two independent photons, Eur. Phys. J. D \textbf{22}, 295 (2003).\\

[32] J. A. Mart\'inez, R. A. Parker, K. C. Chen, . M. Purser, L. Li, C. P. Michaels, A. M. Stramma, R. Debroux, I. B. Harris, M. H. Appel, \textit{et al.}, Photonic Indistinguishability of the Tin-Vacancy Center in Nanostructured Diamond, Phys. Rev. Lett. \textbf{129}, 173603 (2022).\\

[33] R. Proux, M. Maragkou, E. Baudin, C. Voisin, P. Roussignol, and C. Diederichs, Measuring the Photon Coalescence Time Window in the Continuous-Wave Regime for Resonantly Driven Semiconductor Quantum Dots, Phys. Rev. Lett. \textbf{114}, 067401 (2015).\\

[34] J.-H. Kim, T. Cai, C. J. K. Richardson, R. P. Leavitt, and E. Waks, Two-photon interference from a bright single-photon source at telecom wavelengths, Optica \textbf{3}, 577 (2016).\\

[35] C. Nawrath, F. Olbrich, M. Paul, S. L. Portalupi, M. Jetter, and P. Michler, Coherence and indistinguishability of highly pure single photons from non-resonantly and resonantly excited telecom C-band quantum dots, Appl. Phys. Lett. \textbf{115}, 023103 (2019).\\

[36] R. Loudon, The Quantum Theory of Light, 2nd ed. (Oxford
University Press, New York, 1983).\\

[37] K. A. Fischer, K. M\"uller, K. G. Lagoudakis, and J. Vu\v{c}kovi\'c, Dynamical modeling of pulsed two-photon interference, New J. Phys. \textbf{18}, 113053 (2016).\\

[38] J. R. Johansson, P. D. Nation, and F. Nori, QuTiP: An opensource Python framework for the dynamics of open quantum systems, Comput. Phys. Commun. \textbf{183}, 1760 (2012).\\

[39] J. R. Johansson, P. D. Nation, and F. Nori, QuTiP 2: A Python framework for the dynamics of open quantum systems, Comput. Phys. Commun. \textbf{184}, 1234 (2013).
\end{thebibliography}
\end{document}